\documentstyle[prl,aps,epsf,floats,twocolumn]{revtex}

\begin{document}
\draft

\author{Akihisa Koga and Norio Kawakami}
\address{Department of Applied Physics, 
Osaka University, Suita, Osaka 565-0871, Japan}
\title{Quantum Phase Transitions
in the Shastry-Sutherland Model for $\rm\bf SrCu_2 (BO_3)_2$}

\date{December 6, 1999}
\sloppy

\wideabs{

\maketitle

\begin{abstract} 
We investigate the quantum phase transitions in the frustrated
antiferromagnetic Heisenberg model for $\rm SrCu_2(BO_3)_2$
by using the series expansion method. It is found that a novel 
spin-gap phase, which is adiabatically connected to 
the plaquette-singlet phase,
exists between the dimer and the magnetically ordered phases 
known so far. When the ratio of the competing exchange 
couplings $\alpha(=J'/J)$  is varied, this spin-gap phase exhibits
the first- (second-) order quantum phase transition to the dimer 
(the magnetically ordered) phase at the critical point 
$\alpha_{c1}=0.677(2)$ ($\alpha_{c2}=0.86(1)$). Our results shed 
light on some controversial arguments about the nature of the 
quantum phase transitions in this model.\\
\end{abstract}



}

\narrowtext

\begin{figure}[htb]
\epsfxsize=7cm
\centerline{\epsfbox{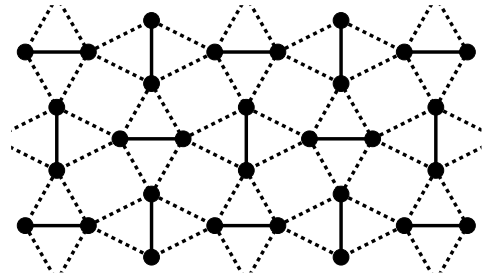}} 
\vspace{0.1cm}
\caption{Lattice structure of the $\rm Cu^{2+}$ spins of $\rm SrCu_2(BO_3)_2$.
The nearest-neighbor bonds ($J$) are expressed by the solid lines and 
the next-nearest-neighbor bonds ($J')$ by the dashed lines.}
\label{fig:lattice}
\end{figure}

Two-dimensional (2D) antiferromagnetic quantum spin systems with 
the spin gap have been the subject of considerable interest.
A typical compound  found recently is 
$\rm SrCu_2(BO_3)_2$\cite{Kageyama}, in which the characteristic 
lattice structure of the $\rm Cu^{2+}$ 
spins (see Fig. \ref{fig:lattice})
stabilizes the singlet ground state.
This system has been providing a variety of interesting phenomena
such as the plateaus in the magnetization curve 
observed at $1/3, 1/4$ and $1/8$ of the full moment\cite{Kageyama,Onizuka}.
The spin system may 
be described by the 2D Heisenberg model on the square lattice with 
some diagonal bonds which is referred to as 
the Shastry-Sutherland model\cite{Shastry},
as pointed out by Miyahara and Ueda\cite{Miyahara}.
The key structure with the orthogonal dimers shown in Fig. \ref{fig:lattice} 
makes the system unique and particularly interesting 
among 2D spin-gap compounds.
In this frustrated system, there may occur 
non-trivial quantum phase transitions when the nearest-neighbor coupling 
$J$ and the  next-nearest-neighbor coupling $J'$ are varied.
Albrecht and Mila\cite{Mila} discussed the possibility of
a helical phase between 
the dimer and the magnetically ordered phases 
by means of the Schwinger boson mean-field theory.
Recent theoretical studies, however, 
have suggested that there may not be such a helical phase, but
the first-order phase transition  occurs from the 
dimer to the ordered phases \cite{Miyahara,Weihong}.
Furthermore, more recent study\cite{Muller} 
claims that the phase transition should be of the second order 
with a non-trivial critical exponent $\nu=0.45(2)$.
These controversial conclusions may come from the fact that 
the quantum phase transition in the Shastry-Sutherland model
suffers from the strong frustration
due to the competing exchange interactions $J$ and $J'$,
and therefore a careful treatment should be necessary to 
figure out the correct nature of the phase transition.
 In particular, we  have to keep in mind 
that such a strong frustration may possibly stabilize another 
spin-gap phase distinct from the dimer phase.

In this paper, by calculating the ground state energy, 
the staggered susceptibility and the spin gap
by means of the series expansion method,
we find  that there should exist a novel spin-gap phase 
with the  disordered ground state,
which is stabilized  by the strong frustration,
between the dimer and the magnetically ordered phases.
The spin-gap phase found in this paper undergoes 
the first- (second-) order quantum phase transition to the dimer 
(the ordered) phase, when the 
exchange couplings $J$ and $J'$ are varied.
The existence of the new phase can resolve controversial 
conclusions \cite{Miyahara,Mila,Weihong,Muller} deduced for 
the quantum phase transitions in this frustrated model.
We also point out that the material  $\rm SrCu_2(BO_3)_2$ lies 
around the phase boundary between these two spin-gap phases,
which may give a natural interpretation  for 
the 1/8-plateau formation in the magnetization curve.

\begin{figure}[htb]
\epsfxsize=7cm
\centerline{\epsfbox{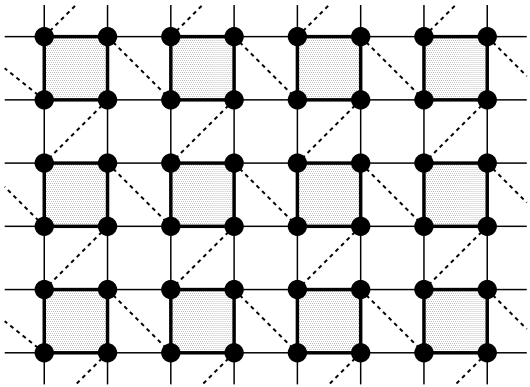}} 
\vspace{0.1cm}
\caption{2D spin system with the plaquette structure. 
The solid circle represents the $s=1/2$ spin.
The bold solid, the thin solid and the dashed lines represent 
the coupling constants $J'$, $\lambda J'$ and $\lambda J'/\alpha$.
When $\lambda=1$, this system is reduced to the Shastry-Sutherland model
for $\rm SrCu_2(BO_3)_2$.}
\label{fig:model}
\end{figure}

To investigate  the frustrated spin system for
the compound $\rm SrCu_2(BO_3)_2$,
we consider the 2D quantum Heisenberg model 
(Shastry-Sutherland model\cite{Shastry,Miyahara}),
which is described by the following Hamiltonian 
\begin{eqnarray}
H=J\sum_{n.n.}{\bf S}_i\cdot{\bf S}_j+J'\sum_{n.n.n.}{\bf S}_i\cdot{\bf S}_j,
\label{eq:model}
\end{eqnarray}
where ${\bf S}_i$ is the $s=1/2$ spin operator at the $i$-th site and 
$J$ $(J')$ represents the nearest-neighbor (next-nearest-neighbor) 
antiferromagnetic exchange coupling.
For later convenience, we introduce  the ratio $\alpha=J'/J$.
In Fig. \ref{fig:lattice}, we have drawn the 2D Heisenberg model schematically.
We note that the system with only the next-nearest-neighbor coupling $J'$ 
is equivalent to the Heisenberg model on the square lattice 
which has a spontaneous staggered magnetization at $T=0$\cite{CHN,QMC}.
From  this point of view, the nearest-neighbor coupling $J$ is
regarded as the coupling for a diagonal bond (see Fig. \ref{fig:model}),
which gives rise to the frustration together with $J'$
\cite{Shastry,Miyahara}.

In order to study the quantum phase transitions in this spin system,
we employ the series expansion method developed by 
Singh, Gelfand and Huse\cite{first}.
We recall here that the quantum phase 
transitions  in the Shastry-Sutherland model have been discussed 
by Weihong et al.\cite{Weihong} and M\"uller-Hartmann et al.\cite{Muller},
by means of  the dimer and the Ising expansions,
from which the critical point between the dimer phase 
and the magnetically ordered phase
has been estimated as $\alpha_{c}=0.691(6)$ and $0.697(2)$, 
respectively. As mentioned above, however,  there is a
controversy to be resolved about the nature of the phase transitions. 
Also, in order to determine the complete phase diagram,
it is crucial to figure out whether there
may exist another spin-gap phase besides the above two phases.
We will address this problem in the following by
using the series expansion method.

To see our strategy clearly, we start with the 2D quantum spin model 
schematically shown in Fig. \ref{fig:model} \cite{Shastry,Miyahara}, which 
is topologically equivalent to the original model in Fig. \ref{fig:lattice}.
In this figure, we have introduced an auxiliary
parameter $\lambda$, which parameterizes
the antiferromagnetic couplings labeled by
the bold solid, the thin solid and the dashed lines, respectively, 
as $J'$, $\lambda J'$ and $\lambda J'/\alpha (=\lambda J)$.  Note that
the system is reduced to the original Shastry-Sutherland model 
in the case of $\lambda=1$. An important point is that 
the introduction of $\lambda$ enables us 
to  perform the cluster expansion starting from the isolated 
{\it plaquette singlets } ($\lambda=0$), which naturally describes
the most likely spin-gap phase distinct from the dimer phase.

To proceed the analysis based on the series expansion,
we divide the original Hamiltonian eq. (\ref{eq:model}) 
into two parts as $H=J' \left[ \sum{\bf S}_{i}\cdot{\bf S}_{j}+
\lambda\sum\Gamma_{ij}{\bf S}_{i}\cdot{\bf S}_{j}\right]$,
where $\Gamma_{ij}=1$ or $\alpha^{-1}$ for each bond on the square lattice
(see Fig. \ref{fig:model}).
The first term is the unperturbed Hamiltonian 
which stabilizes the isolated plaquette singlets with the
spin excitation gap.
The perturbed  Hamiltonian labeled by $\lambda$
connects these isolated plaquette singlets, 
by which a 2D network develops.
We expand the staggered susceptibility $\chi_{\rm AF}$, 
the spin-triplet excitation energy  $E({\bf k})$ 
and the ground state energy $E_{\rm g}$ as a power series in $\lambda$.
Here, to estimate the susceptibility, 
we introduce the Zeeman term 
$H'=h\left[\sum_{i\in A}S^z_i-\sum_{i\in B}S^z_i\right]$,
where $h$ is the staggered magnetic field and $A(B)$ denotes 
one of the two sublattices. 
Note that an asymptotic analysis of the series expansion
is necessary to deduce the accurate phase boundary on which 
the susceptibility $\chi_{\rm AF}$ diverges and 
the spin gap $\Delta=E({\bf k}={\bf 0})$ vanishes.
For this purpose, we make use of the Pad\'e approximants\cite{Pade} 
for both  quantities obtained  up to the finite order in $\lambda$.
Besides ordinary Dlog Pad\'e 
approximants, we also employ {\it biased} Pad\'e approximants\cite{Pade},
for which we assume that the phase transition in our 2D quantum spin models 
should belong to the universality class of the 3D 
classical Heisenberg model\cite{CHN}. 
Then the critical value of $\lambda_{c}$ 
is determined by the formula 
$\chi_{\rm AF} \sim (\lambda_{c}-\lambda)^{-\gamma}$ and
$\Delta \sim (\lambda_{c}-\lambda)^{\nu}$
with the known exponents $\gamma=1.4$ and $\nu=0.71$\cite{Ferer}. 

\begin{figure}[htb]
\epsfxsize=8cm
\centerline{\epsfbox{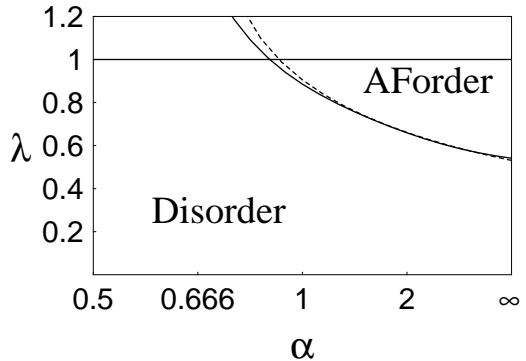}} 
\vspace{0.1cm}
\caption{Phase diagram for the 2D spin system with the plaquette structure 
in Fig. \protect{\ref{fig:model}}.
The solid (dashed) line indicates the phase boundary obtained by 
biased Pad\'e approximants for the spin gap 
(the staggered susceptibility). 
}
\label{fig:diagram}
\end{figure}

We first calculate the staggered susceptibility $\chi_{\rm AF}$ and 
the spin gap $\Delta$ by means of the plaquette expansion 
up to the fourth and the fifth order in $\lambda$, respectively,  
for various values of $\alpha$.
Using the Dlog and the biased Pad\'e approximants, 
we end up with  the phase diagram shown 
in Fig. \ref{fig:diagram}.
In this figure, the solid (dashed) line represents the phase boundary 
obtained by the biased Pad\'e approximants 
for the spin gap (the staggered susceptibility).
When $\alpha\rightarrow\infty$ and $\lambda=0$, 
the system is reduced to an assembly of the isolated plaquettes
with the spin gap.  As  $\lambda$ is increased, 
the correlation between these plaquettes grows up and
the second-order quantum phase transition from the spin-gap phase
to the magnetically ordered phase occurs  at the critical 
point $\lambda_{c}=0.56$ for $\alpha \rightarrow \infty $,
which has already been studied by several 
groups\cite{Fukumoto,pladis,plakoga}.
On the other hand, decreasing $\alpha$ enhances the frustration, 
which in turn suppresses
the antiferromagnetic correlation, thus shifting 
the phase boundary upward for smaller $\alpha$ in the phase diagram.
It is seen that two lines  obtained from the distinct quantities 
are in good agreement with each other, which implies that
the obtained  phase boundary is rather accurate
in spite of the lower-order  pertubative calculation.
By exploiting the phase boundary determined 
by means of biased Pad\'e approximants for the spin gap,
the critical value is given by $\alpha_{c2}=0.86(1)$ for $\lambda=1$.
Recall that the system is reduced to the original  model
 only for  $\lambda=1$.
We thus find that the Shastry-Sutherland model 
 has the disordered ground state 
in the region ($0< \alpha < \alpha_{c2}$) on the $\lambda=1$ line.

\begin{figure}[htb]
\epsfxsize=8cm
\centerline{\epsfbox{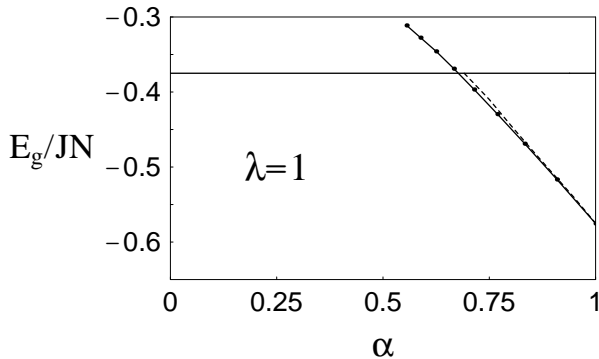}} 
\vspace{0.1cm}
\caption{Ground state energy per site as a function of $\alpha=J'/J$
($\lambda=1$, Shastry-Sutherland model).
The flat line ($E_{\rm g}/JN=-3/8$) is the energy of the exact dimer state, 
while the solid line with dots 
(error bars are smaller than the line width) is 
obtained by  the plaquette expansion.  For comparison, we also show  
the ground state energy obtained 
by Ising expansion\protect{\cite{Weihong}} as the dashed line. 
}
\label{fig:eg}
\end{figure}

\begin{figure}[htb]
\epsfxsize=8cm
\centerline{\epsfbox{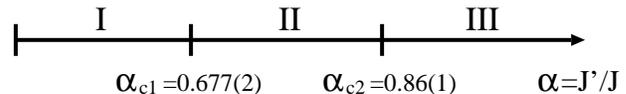}} 
\vspace{0.1cm}
\caption{Phase diagram for the Shastry-Sutherland model.
The phase I represents the orthogonal dimer phase. 
The phase II newly obtained is adiabatically connected to
the plaquette singlet phase.
III is the magnetically ordered phase.}
\label{fig:phase}
\end{figure}

The above result does not necessarily imply that 
in the region  $0< \alpha < \alpha_{c2}$ 
the system always belongs to the disordered phase
which is continuously connected to  isolated plaquettes.
In fact, it is known that the orthogonal dimer phase appears 
in the vicinity of $\alpha=0$ \cite{Shastry,Miyahara}. 
Therefore, it is necessary to clarify how these two spin-gap 
phases compete with each other by carefully comparing the
ground state energy $E_{\rm g}$.
To this end, performing the plaquette expansion up to the seventh order 
in $\lambda$ with $\alpha$ being fixed,
we estimate the ground state energy $E_{\rm g}$ 
for the Shastry-Sutherland model $(\lambda=1)$ 
by means of the first order inhomogeneous differential method\cite{Pade}.
The results are  shown in Fig. \ref{fig:eg}.
As mentioned above\cite{Shastry,Miyahara}, the system 
stabilizes  the  orthogonal dimer ground state for 
smaller $\alpha$.   It is found,
however, that the first-order transition to the novel spin-gap phase 
introduced here occurs at the critical point $\alpha_{c1}=0.677(2)$.
It is also seen from this figure that  further 
increase of $\alpha$ induces the  antiferromagnetic order, 
whose transition point is determined by the crossing point
of the ground-state energy obtained respectively by the Ising\cite{Weihong} 
and plaquette expansions. The result confirms the second-order
phase transition deduced  above, and the transition point estimated from the 
figure  is consistent with  $\alpha_{c2}=0.86(1)$ obtained 
by the analysis of the susceptibility and the spin gap.
Consequently, we end up with the phase diagram 
for the Shastry-Sutherland model as shown in Fig. \ref{fig:phase}.
The present results shed light on the controversial 
arguments whether the quantum phase transition in this 
model is of the first or second order\cite{Miyahara,Weihong,Muller}.
In those previous studies, it was believed that the phase transition 
occurs only once between the dimer phase (I) and the ordered phase (III), 
giving rise to some confusions.  Our phase diagram clearly resolves
this  problem by explicitly showing the existence of
the new spin-gap phase (II) which undergoes the
first- (I$\leftrightarrow$II) as well as the second-order transitions 
(II$\leftrightarrow$III).

\begin{figure}[htb]
\epsfxsize=8cm
\centerline{\epsfbox{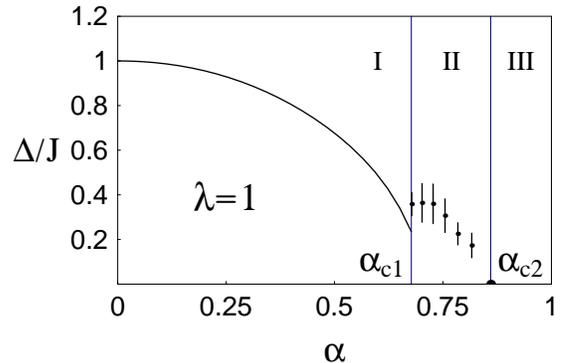}} 
\vspace{0.1cm}
\caption{The spin gap as a function of $\alpha=J'/J$ 
for the Shastry-Sutherland model.
The solid line for $\alpha<\alpha_{c1}$ is the result 
obtained by Weihong et al. \protect{\cite{Weihong}} 
The dots with error bars for $\alpha_{c1}<\alpha<\alpha_{c2}$ 
represent the spin gap at ${\bf k}={\bf 0}$ 
obtained by the plaquette expansion.
}
\label{fig:gap}
\end{figure}

To check the validity of the above phase diagram, 
we also show the results for the spin gap as a function 
of $\alpha=J'/J$ in Fig. \ref{fig:gap}.
In this figure, the results obtained by Weihong et al.\cite{Weihong} 
are shown for the orthogonal dimer phase (I: $0<\alpha<\alpha_{c1}$).
In the new phase (II: $\alpha_{c1}<\alpha<\alpha_{c2}$), 
we determine the values of the spin gap at ${\bf k}={\bf 0}$
by means of the plaquette expansion up to the fifth order in $\lambda$ 
with the first order inhomogeneous differential method. 
The results are shown as the dots with the error bars. 
As seen in this figure, with the decrease of $\alpha$ from 
the second-order transition point $\alpha_{c2}$, 
the spin-gap continuously grows up to stabilize the disordered ground state.  
As $\alpha$ is further decreased, 
the first-order phase transition occurs at $\alpha_{c1}$.  


In order to further confirm the  present results,
we have performed a different series expansion by choosing 
the isolated plaquettes with diagonal bonds
as an initial configuration,
which is different from the one shown in Fig. \ref{fig:model}.
The calculation of the susceptibility up to the fourth order 
yields second-order transition with  
$\alpha_{c2}=0.87(3)$, being consistent with the above results.
Furthermore, to confirm  the first-order 
phase transition between the two spin-gap states, we have checked how
the first-order phase transition point known for the 1D
orthogonal-dimer chain \cite{Ivanov} evolves with the increase of
the inter-chain couplings. By performing the exact diagonalization 
studies for the  $4\times4$ system,
 we have found that the first-order transition 
point for 1D is continuously changed, and in the 
Shastry-Sutherland case, it coincides with the one found
above within reasonable accuracy $(\alpha_{c1}\sim0.66)$, providing
further support to our conclusion on the phase diagram.
Although our results still seem to
be partly contradicted to the staggered magnetization
obtained by Weihong et al.\cite{Weihong}, we believe that this could be 
resolved by further analysis of the results of the Ising 
expansion.

Before concluding the paper, a brief comment is in order for the 
plateau-formation in the magnetization curve.
Experimentally, the plateaus in the magnetization curve
have been observed  for the compound $\rm SrCu_2(BO_3)_2$
at 1/3, 1/4 and $1/8$ of the full moment\cite{Kageyama,Onizuka}.
In the theoretical studies\cite{Miyapla,Momoi,FMag} on the dimer phase,
it has been clarified that the stripe order of the isolated 
dimer-triplets is important to understand 
the $1/3-$ and $1/4-$plateaus. On the other hand, it is not so trivial why 
the 1/8-plateau occurs in this compound, 
although a possible mechanism has been 
proposed \cite{Miyapla,Momoi}.  We think that the formation of 
the 1/8 plateau may reflect the fact that this compound is
located around the first-order phase transition point between 
the two spin-gap phases and thereby possesses 
the dual properties inherent in two distinct phases
implicitly. We note here  that the new spin-gap phase 
belongs to the same phase
as the Heisenberg model on the 1/5-depleted square lattice proposed
for $\rm CaV_4O_9$.
Therefore it is likely that the 1/8-plateau could occur in the
same origin discussed by Momoi and Totsuka\cite{Momoi}
for the plaquette system related to the 1/5-depleted Heisenberg model.
It is interesting to further clarify the mechanism of the 1/8-plateau
by taking into account the above dual properties  explicitly,
which is now under consideration.

In conclusion we have discussed the phase diagram
for the Shastry-Sutherland model for the compound $\rm SrCu_2(BO_3)_2$
by means of the series expansion method.
Our analysis has shown that there exists
a novel spin-gap phase with the disordered ground state, which is 
adiabatically connected to the plaquette-singlet phase,
between the dimer and the magnetically ordered phases known so far.
When the exchange coupling ratio $\alpha=J'/J$ is varied,
the first-order phase transition occurs from the dimer state 
to the new spin-gap state, while  
the second-order phase transition occurs from this spin-gap
state to the magnetically ordered state.
This sheds light on the  nature of the quantum phase 
transitions in this model,
and resolves apparently controversial conclusions on this issue.

We would like to thank K. Ueda, S. Miyahara and  K. Okunishi 
for useful discussions. The work is partly supported by a 
Grant-in-Aid from the Ministry of Education, Science, Sports, 
and Culture. A. K. is supported by the Japan Society 
for the Promotion of Science. 
A part of numerical computations in this work was carried out 
at the Yukawa Institute Computer Facility.


\begin{thebibliography}{99}

\bibitem{Kageyama}
H. Kageyama, K. Yoshimura, R. Stern, N. V. Mushnikov, K. Onizuka, M. Kato,
K. Kosuge, C. P. Slichter, T. Goto and Y. Ueda,
Phys. Rev. Lett. {\bf 82}, 3168 (1999).

\bibitem{Onizuka}
K. Onizuka, H. Kageyama, Y. Narumi, K. Kindo, Y. Ueda and T. Goto,
preprint.

\bibitem{Shastry}
B. S. Shastry and B. Sutherland, Physica {\bf 108B}, 1069 (1981).

\bibitem{Miyahara}
S. Miyahara and K. Ueda, Phys. Rev. Lett. {\bf 82}, 3701 (1999).

\bibitem{Mila}
M. Albrecht and F. Mila, Europhys. Lett. {\bf 34}, 145 (1996).

\bibitem{Weihong}
Z. Weihong, C. J. Hamer and J. Oitmaa, Phys. Rev. B {\bf 60}, 6608 (1999).

\bibitem{Muller}
E. M\"uller-Hartmann, R. R. P. Singh, C. Knetter and G. S. Uhrig,
cond-mat / 9910165.

\bibitem{CHN}
S. Chakravarty, B. I. Halperin and D. R. Nelson,
Phys. Rev. B {\bf 39}, 2344 (1989).

\bibitem{QMC}
J. D. Reger and A. P. Young, Phys. Rev. B {\bf 37}, 5978 (1988).

\bibitem{first}
R. R. P. Singh, M. P. Gelfand and D. A. Huse,
Phys. Rev. Lett. {\bf 61}, 2484 (1988). 




\bibitem{Pade}
A. J. Guttmann, in {\it Phase Transitions and Critical Phenomena},
edited by C. Domb and J. L. Lebowitz 
(Academic, New York, 1989), Vol. 13.


\bibitem{Ferer}
M. Ferer and A. Hamid-Aidinejad,
Phys. Rev. B {\bf 34}, 6481 (1986).
\bibitem{Fukumoto}
Y. Fukumoto and A. Oguchi,
J. Phys. Soc. Jpn. {\bf 67}, 697 (1998); 
J. Phys. Soc. Jpn. {\bf 67}, 2205 (1998).

\bibitem{pladis}
Z. Weihong, J. Oitmaa and C. J. Hamer,
Phys. Rev. B {\bf 58}, 14147 (1998);
R. R. P. Singh, Z. Weihong, C. J. Hamer and J. Oitmaa,
cond-mat / 9904064.

\bibitem{plakoga}
A. Koga, S. Kumada and N. Kawakami,
J. Phys. Soc. Jpn. {\bf 68}, 2373 (1999); cond-mat / 9908458.

\bibitem{Ivanov}
N. B. Ivanov and J. Richter,
Phys. Lett. {\bf 232A}, 308 (1997).

\bibitem{Miyapla}
S. Miyahara and K. Ueda, preprint.

\bibitem{Momoi}
T. Momoi and K. Totsuka, cond-mat / 9910057.

\bibitem{FMag}
Y. Fukumoto and A. Oguchi, preprint.






\end{thebibliography}
\end{document}